# Quantum Entanglement with Self-stabilizing Token Ring for Fault-tolerant Distributed Quantum Computing System

Jehn-Ruey Jiang

Department of Computer Science and Information Engineering
National Central University
Taoyuan City, 23001, Taiwan

**Abstract**

This paper shows how to construct quantum entanglement states of *n* qubits based on a self-stabilizing token ring algorithm. The entangled states can be applied to the fields of the quantum network, quantum Internet, distributed quantum computing, and quantum cloud. To the best of our knowledge, this is the first attempt to construct quantum entanglement based on the self-stabilizing algorithm. By the quantum circuit implementation based on the IBM Quantum Experience platform, it is demonstrated that the construction indeed can achieve specific *n*-qubit entangled states, which in turn can be used to circulate a token in a quantum network or quantum Internet for building a distributed quantum computing system (DQCS). The built DQCS is fault-tolerant in the sense that it can tolerate transient faults such as occasional errors of entangled quantum states.

**Keywords:** quantum entanglement, self-stabilizing system, token ring, fault-tolerance, distributed quantum computing system, quantum network, quantum Internet

## 1. Introduction

Quantum entanglement, introduced by Schrodinger [1] and referred to as "spooky action at a distance" by Einstein, is very critical in quantum mechanics. It is the physical phenomenon of a group of quantum entities which interact with one another. The entangled quantum entities' properties have been integrated into an overall property. Thus, when the state of an entity changes, the states of other entities change instantaneously no matter how far the entangled entities are. For example, regarding two entangled photons with opposite polarization, if one photon is observed (or measured) to be vertically polarized, then the other photon must be observed to be horizontally polarized. Quantum entanglement has been applied to many applications, such as the quantum key distribution with entanglement [2], quantum teleportation [3, 4], quantum positioning systems [5], quantum approximate optimization algorithms [6, 7], quantum networks [8], quantum Internet [9], distributed quantum computing [10], and quantum cloud [10].

Bhatia et al. [11] modeled quantum entities as qubits to implement their entangled states. The implemented entangled states are the Greenberger-Horne-Zeilinger (GHZ) state [12] and the W state [13] for *n* qubits. In the GHZ state, either all *n* qubits are of the state $|0\rangle$ or all are of the state $|0\rangle$. In the W state, one of the *n* qubits is of the state $|1\rangle$, whereas other qubits are of the state $|0\rangle$. Experiments based on the IBM Quantum Experience platform are conducted to show probability distribution of qubit states to validate the implementation correctness.

This paper proposes constructing quantum entanglement states of *n* qubits based on a self-stabilizing token ring algorithm [14]. The entangled states can be applied to the fields of the quantum network, quantum Internet, distributed quantum computing, and quantum cloud. To the best of our knowledge, this is the first attempt to realize quantum entanglement based on the self-stabilizing algorithm. It is well known that self-stabilizing algorithms are fault-tolerant in the sense that they can tolerant transient faults. As will be discussed later, the quantum entanglement state realized in this paper can be applied to a distributed quantum computing system (DQCS) to make it fault-tolerant.

A self-stabilizing algorithm can realize a self-stabilizing system. A system is said to be self-stabilizing if it satisfies the following two properties: (1) Convergence: Starting from any initial state, the system can converge to a legitimate state in finite time. (2) Closure: When the system is in one of the legitimate states, it remains so henceforth. The concept of the self-stabilizing system was proposed in1974 by Dijkstra [15]. It can be used to enable a distributed computing system (DCS) consisting of interconnected state machines to tolerate transient faults, that is, occasional machine state readout errors. The quantum entanglement states constructed in this paper includes all legitimate states of the self-stabilizing system. They can be applied to build a fault-tolerant DQCS that can tolerant transient quantum entanglement state errors.

The rest of this paper is organized as follows. Section 2 introduces some background knowledge. Section 3 shows the quantum circuit implementation of the self-stabilizing token ring algorithm based on the IBM Quantum Experience platform. The probability distribution of qubits is also shown in this section for verifying that the qubits are indeed in quantum entanglement. Finally, some concluding remarks are drawn in section 4.

## 2. Preliminaries

### 2.1 Quantum entanglement of $n$ qubits

This subsection introduces two well-known entangled states of $n$ qubits, the Greenberger-Horne-Zeilinger (GHZ) state [12] and the W state [13], and their implementation [11] based on the IBM Quantum Experience platform.

- GHZ quantum entanglement:

The GHZ quantum entanglement state of $n$ qubits is defined as

$$|GHZ\rangle = \frac{1}{\sqrt{2}}(|0\rangle^{\otimes n} + |1\rangle^{\otimes n})$$

The quantum circuit to realize the GHZ quantum entanglement state of 3 qubits and its probability distribution are shown in Figure 1.

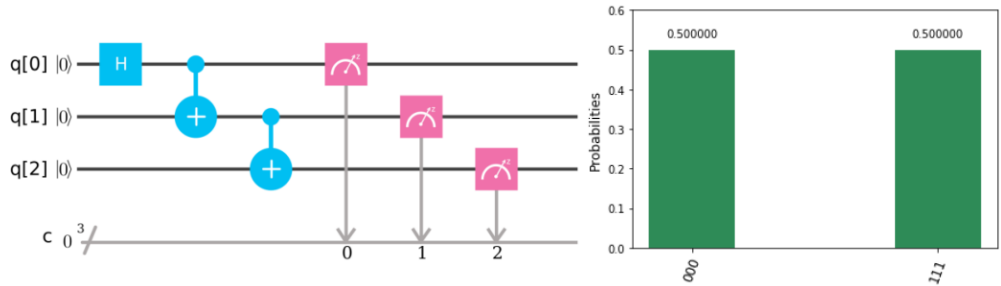

Figure 1. The quantum circuit for 3-qubit GHZ quantum entanglement state and its probability distribution [11].

- W quantum entanglement:

The W quantum entanglement state of $n$ qubits is defined as

$$|W\rangle = \frac{1}{\sqrt{n}}(|10\ldots0\rangle + |01\ldots0\rangle+\ldots+|00\ldots1\rangle)$$

The quantum circuit to realize the W quantum entanglement state of 3 qubits and its probability distribution are shown in Figure 2.

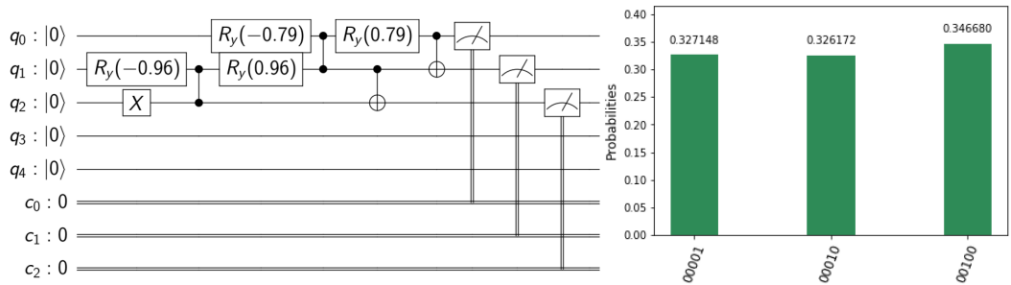

Figure 2. The quantum circuit for 3-qubit W quantum entanglement state and its probability distribution [11].

## 2.2 Distributed Quantum Computing System

Today's quantum processors can operate on many qubits simultaneously, and their computing power grows exponentially with the number of qubits. However, increasing the number of qubits on a single quantum processor is extremely challenging. Combining two isolated quantum processors just increases the computing power by a factor of 2. However, clustering (or interconnecting) two quantum processors through quantum entanglement can keep the computing power to increase exponentially with the number of interconnected qubits. For example, a 10-qubit quantum processor can present and manipulate $2^{10}$ quantum states at a time due to the superposition property of quantum mechanics. Combining two isolated 10-qubit quantum processors can present and manipulate $2 \times 2^{10}$ quantum states. However, clustering (or interconnecting) two 10-qubit quantum processors can then present and manipulate $2^{18}$ states at a time by using one qubit of each quantum processor to form quantum entanglement for exchanging qubit states between the two processors. Figure 3 shows the speed-up of combining isolated quantum processors (or quantum devices) and clustering entangled quantum processors [10].

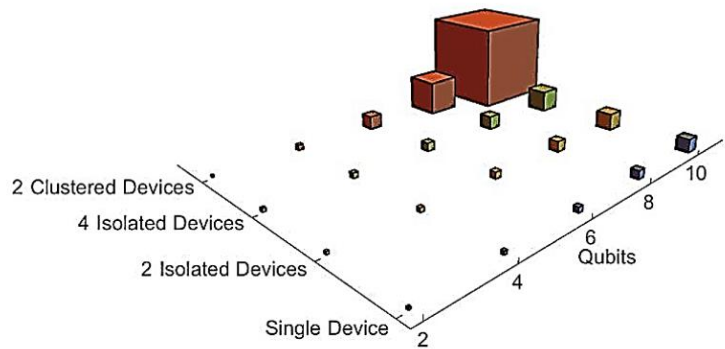

Figure 3. Distributed quantum computing system speed-up (the volume of cubes stands for the quantum computing power without noise or errors) [10].

It is a hot research topic to form a distributed quantum computing system (DQCS) interconnecting quantum processors with quantum entanglement through quantum networks such as the quantum Internet. Quantum entanglement is the core of the quantum Internet. It can be relied upon to transmit qubit states without violating the non-cloning theorem. Using only local operations and a pair of entangled qubits shared between the source and the destination, an unknown quantum state can be transmitted between two remote quantum processors or devices. A DQCS can be regarded as a virtual quantum computer having control of all qubits of connected quantum processors. It is very promising since its computing power still grows exponentially with the total number of qubits of quantum processors connected to the DQCS.

To execute a quantum task on the DQCS requires a precise execution schedule. Moreover, when two quantum processors are too far away, the transmission of qubit states may go erroneously. Therefore, the self-stabilizing algorithm that can tolerate transient faults can be introduced into the DQCS for the sake of distributed autonomous control with the fault-tolerant property.

**2.3 A self-stabilizing token ring algorithm**

In 1974, Dijkstra introduced the concept of self-stabilization [15] for traditional distributed computing systems (DCSs). A DCS can be considered as a network of classical processors or nodes keeping one or more variables. Each node performs specific actions based on the values of its own variables and their neighboring nodes. As mentioned earlier, a DCS is said to be self-stabilizing if it satisfies the following two properties: (1) Convergence: Starting from any initial state, the system can converge to a legitimate state in finite time. (2) Closure: When the system is in one of the legitimate states, it remains so henceforward. That is, a self-stabilizing DCS guarantees to converge to a legitimate state in finite time no matter what the initial state is. From time to time, a transient fault occurs to make a node read the variable value of its neighboring node erroneously, which can be regarded as if the DCS restarts with an initial illegitimate state. By the convergence and the closure properties, the system will recover to legitimate state autonomously. Therefore, a self-stabilizing DCS is fault-tolerant in the sense that it can tolerate transient faults.

A large number of self-stabilizing algorithms have been proposed in the literature. They are algorithms to keep DCSs in legitimate states for accomplishing certain tasks, such as coloring nodes with various colors, constructing trees spanning all nodes, finding maximal matching of nodes, and circulating a token among nodes. Please refer to survey papers [16-18] for their details.

A self-stabilizing algorithm consist of rules for every node, where a rule is of the form:

*guard* → *move*

A guard is a Boolean expression of the variables that can be read by a node (i.e., its own variables and its neighboring nodes'). If a node has a rule whose guard is true, then it is called a privileged node; it has the privilege to make a move. It is possible that many privileged nodes exist at the same time. Different orders of making moves by different nodes take the whole system into different configurations. Theoretically, there are mainly three models for selecting privileged nodes to make moves. The models are the central demon, randomized central demon, and distributed demon.

This paper focuses on the simplest model, the central demon, as described below. Rules are executed atomically, one at a time. The central demon arbitrarily selects one of the privileged nodes to make a move at a time. After the selected node completes the move, the demon re-selects a privileged node, if any, to make another move, and so on. Therefore, the moves of the nodes are serialized; this is why the central demon model is also called the serial execution model. However, the central demon selects privileged nodes in an unpredictable manner, so the execution sequence can be in any possible order.

Flatebo et al. proposed a two-state self-stabilizing algorithm [14] to circulate a token on a token ring network. There are $n$ nodes forming a ring, where each node keeps a two-state variable $x_i$, $0 \leq i \leq n-1$. Each node has only one rule to execute:

Rule: $\begin{cases} x_0 = x_{n-1} \rightarrow x_0 := x_0' \\ x_i \neq x_{i-1} \rightarrow x_i := x_i' \text{ for } i = 1, \dots, n-1 \end{cases}$

The algorithm is shown in [14] to satisfy the convergence property and the closure property under the central demon model to reach the legitimate state that only one node is privileged. The algorithm is also proved to be able to circulate a token in a ring of nodes by regarding the privileged node as the node owning the token. In practice, it is a self-stabilizing algorithm that can tolerate transient faults.

## 3. Proposed Scheme

The self-stabilizing token ring algorithm proposed in [14] enables traditional DCSs to tolerate transient faults, such as occasional machine state readout errors. This paper proposes using the self-stabilizing token ring algorithm [14] under the central demon mode to achieve specific quantum entanglement states of $n$ qubits. If the quantum entanglement states are employed to form quantum networks for building DQCSs, then the DQCSs are fault-tolerant and can tolerate transient faults like quantum entanglement state errors.

By taking $n$ as 3, the rules for nodes $n_0$, $n_1$, and $n_2$ are implemented with quantum circuits based on the IBM Quantum Experience platform. Since the central demon is assumed, the execution of the rule is serialized. The left parts of Figures 4-6 show the quantum circuits corresponding to rules of nodes $n_0$, $n_1$, and $n_2$, respectively. The quantum circuits have three input qubits $x_0$, $x_1$, and $x_2$, along with nine ancilla qubits $a_0$, …, $a_8$, and three classical bits to store the measurement results of qubits $x_0$, $x_1$, and $x_2$. The three input qubits are initialized to be in a uniform superposition state $|\psi\rangle$, as shown below:

$$|\psi\rangle = \frac{1}{\sqrt{8}}(|000\rangle + |001\rangle + |010\rangle + |011\rangle + |100\rangle + |101\rangle + |110\rangle + |111\rangle)$$

The initial state $|\psi\rangle$ consists of all possible eight states of a token ring of three nodes. Some nodes are privileged for a state, whereas some nodes are not privileged for the state. If a node is privileged, it makes a move to flip its state. Therefore, the system is in the state after a certain node has been privileged and made a move. For example, the right part of Figure 4 shows the probability distribution of four entangled quantum states of the self-stabilizing token ring after $n_0$ has been privileged and made a move. The states are thus $|001\rangle$, $|011\rangle$, $|100\rangle$, and $|110\rangle$. The left part of Figure 7 shows the quantum circuit that implements rules of nodes $n_0$, $n_1$, and $n_2$. The right part of Figure 7 shows the probability distribution of two entangled quantum states of the self-stabilizing token ring after $n_0$, $n_1$, and $n_2$ have all been privileged and made a move. The states are thus $|000\rangle$ and $|111\rangle$.

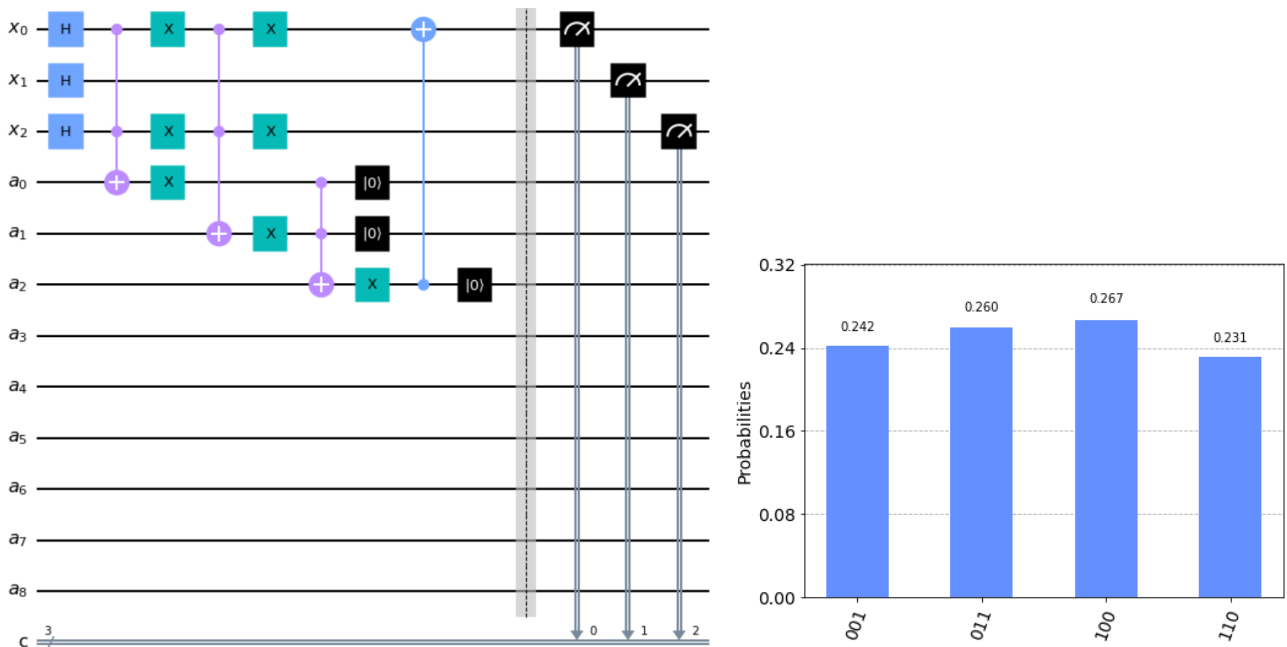

Figure 4. The quantum circuit implementation corresponding to the rule of $n_0$ and the probability distribution of quantum entanglement states after $n_0$ has been privileged and made a move.

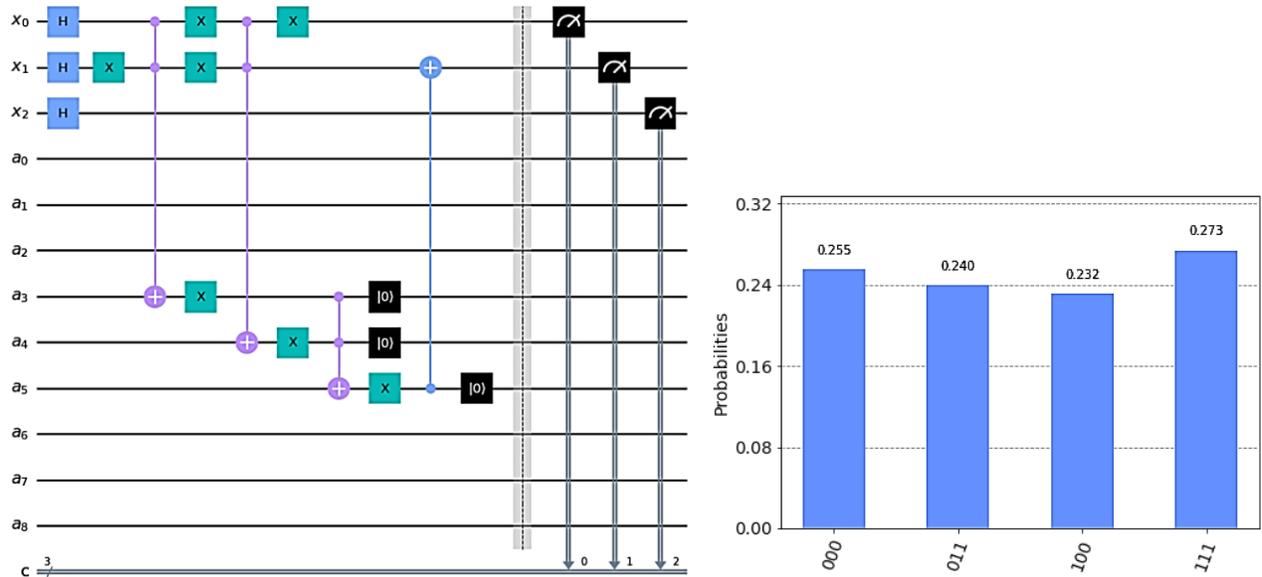

Figure 5. The quantum circuit implementation corresponding to the rule of $n_1$ and the probability distribution of quantum entanglement states after $n_1$ has been privileged and made a move.

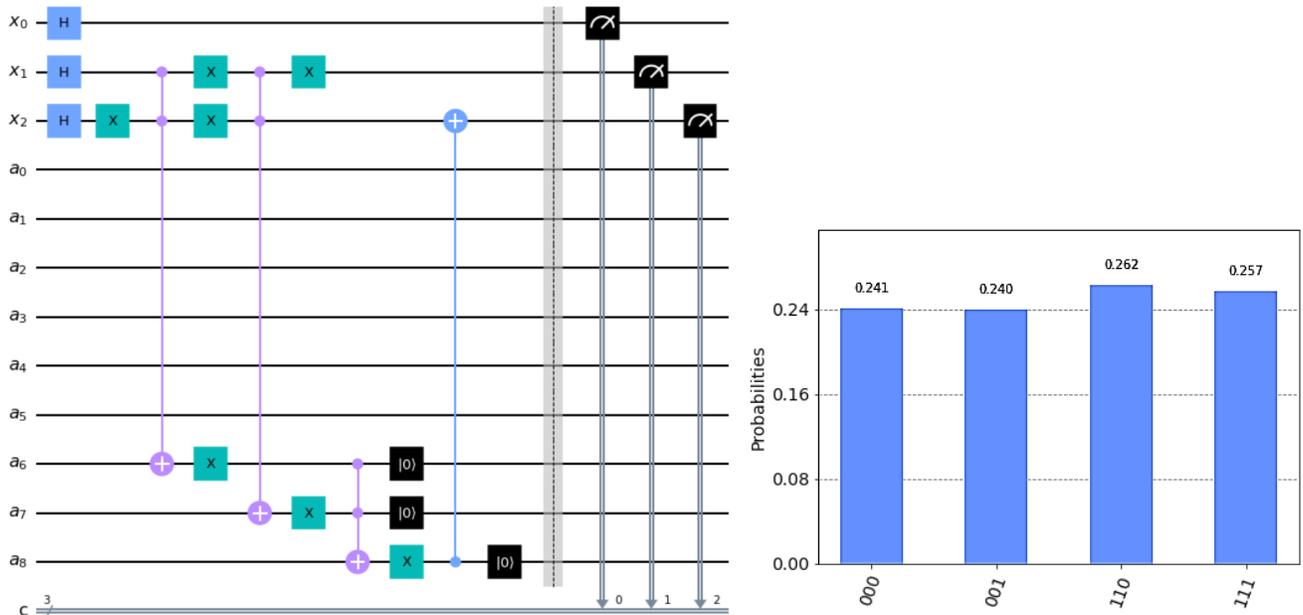

Figure 6. The quantum circuit implementation corresponding to the rule of $n_2$ and the probability distribution of quantum entanglement states after $n_2$ has been privileged and made a move.

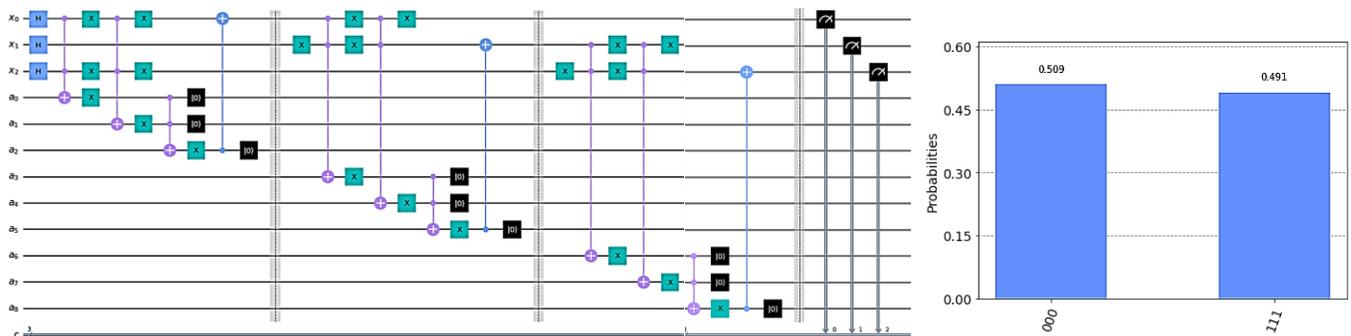

Figure 7. The quantum circuit implementation corresponding to the rules of $n_0$, $n_1$, and $n_2$, and the probability distribution of quantum entanglement states after $n_0$, $n_1$, and $n_2$ have been privileged and made a move.

## 4. Conclusion

This paper proposes constructing specific quantum entanglement of *n* qubits on the basis of a self-stabilizing token ring algorithm [14]. It shows the quantum circuit implementation of the self-stabilizing token ring algorithm based on the IBM Quantum Experience platform. The probability distribution of qubits is also shown for verifying that the qubits are indeed in specific quantum entanglement states corresponding to legitimate states of self-stabilizing algorithms. To the best of our knowledge, this is the first attempt to realize quantum entanglement based on the self-stabilizing algorithm.

The quantum entanglement states realized in this paper can be applied to form the quantum Internet for building a DQCS having fault-tolerant token circulation. In the future, we plan to connect real quantum processors to build DQCSs with the token ring quantum entanglement states for performing token ring-based quantum computation. We also plan to investigate more self-stabilizing algorithms to see if we can construct specific quantum entanglement states based on them.